\documentclass{article}

\usepackage[utf8]{inputenc} %
\usepackage[table]{xcolor}
\usepackage[T1]{fontenc}    %
\usepackage[colorlinks=true,linkcolor=black,anchorcolor=black,citecolor=black,filecolor=black,menucolor=black,runcolor=black,urlcolor=black]{hyperref}       %
\usepackage{url}            %
\usepackage{booktabs}       %
\usepackage{amsfonts}       %
\usepackage{nicefrac}       %
\usepackage{microtype}     %
\usepackage{xcolor}         %
\usepackage{bm}         %
\usepackage{graphicx}  		%
\usepackage{amsmath} %
\usepackage{cite} %
\usepackage{diagbox}
\usepackage{enumitem}
\usepackage{siunitx}
\usepackage{multirow}
\usepackage{tabularx}
\usepackage{hhline}
\usepackage{arydshln}		%
\usepackage{xcolor}
\usepackage{mathtools} %
\usepackage{amsthm}			%
\usepackage{amssymb}			%
\usepackage{algorithmic}	%
\usepackage{algorithm}
\usepackage{pifont}
\usepackage{threeparttable}
\usepackage{fullpage}
\usepackage{caption}
\usepackage{makecell}

\definecolor{lightgreen}{rgb}{.9,1,.9}
\definecolor{lightred}{rgb}{1,.415,.415}
\definecolor{lightblue}{rgb}{.415,.415,1}

\def\argmin{\mathop{\mathrm{arg\,min}}} %

\def\lim{\mathop{\mathrm{lim}}} %
 
\def\max{\mathop{\mathrm{max}}}

\newcommand{\norm}[1]{\left\lVert#1\right\rVert}

\def\cbm{{\bm{c}}}

\def\xbm{{\bm{x}}}

\def\ybm{{\bm{y}}}

\def\vbm{{\bm{v}}}

\def\varphibm{{\bm{\varphi}}}

\def\thetabm{{\bm{\theta}}}

\def\Dbm{{\bm{D}}}
\def\Fbm{{\bm{F}}}

\def\Pbm{{\bm{P}}}

\def\xbmast{{\bm{x}^\ast}}

\def\xbmhat{{\widehat{\bm{x}}}}

\def\xbmbar{{\bar{\xbm}}}

\def\Tsf{{\mathsf{T}}}

\def\Hsf{{\mathsf{H}}}

\def\C{\mathbb{C}}
\def\R{\mathbb{R}}

\def\mbm{{\bm{m}}}

\begin{document}

\markboth{}{Gan {et al.}}

\title{PtychoDV: Vision Transformer-Based Deep Unrolling Network for Ptychographic Image Reconstruction}

\author{
Weijie Gan\textsuperscript{\rm 1}, 
Qiuchen Zhai\textsuperscript{\rm 2}, 
Michael Thompson McCann\textsuperscript{\rm 3}, 
Cristina Garcia Cardona\textsuperscript{\rm 3}, \\
Ulugbek~S.~Kamilov\textsuperscript{\rm 1,4},
Brendt Wohlberg\textsuperscript{\rm 3} \\
\small \textsuperscript{\rm 1}{Department of Computer Science \& Engineering, Washington University in St. Louis, St. Louis, MO, 63110 USA}\\
\small \textsuperscript{\rm 2}{School of Electrical and Computer Engineering, Purdue University, West Lafayette, IN, 47907 USA}\\
\small \textsuperscript{\rm 3}{Theoretical Division, Los Alamos National Laboratory, Los Alamos, NM, 87545 USA}\\
\small \textsuperscript{\rm 4}{Department of Electrical \& System Engineering, Washington University in St. Louis, St. Louis, MO, 63110 USA}\\
}

\date{}

\maketitle

\begin{abstract}
Ptychography is an imaging technique that captures multiple overlapping snapshots of a sample, illuminated coherently by a moving localized probe. The image recovery from ptychographic data is generally achieved via an iterative algorithm that solves a nonlinear phase retrieval problem derived from measured diffraction patterns. However, these iterative approaches have high computational cost. In this paper, we introduce PtychoDV, a novel deep model-based network designed for efficient, high-quality ptychographic image reconstruction. PtychoDV comprises a vision transformer that generates an initial image from the set of raw measurements, taking into consideration their mutual correlations. This is followed by a deep unrolling network that refines the initial image using learnable convolutional priors and the ptychography measurement model. Experimental results on simulated data demonstrate that PtychoDV is capable of outperforming existing deep learning methods for this problem, and significantly reduces computational cost compared to iterative methodologies, while maintaining competitive performance.
\end{abstract}

\section{Introduction}
Ptychography is an essential imaging technique {applied in fields} such as materials science, biology, and nanotechnology, due to its ability to provide high-resolution images of samples~\cite{Pfeiffer2018}. In ptychographic imaging, a localized coherent scanning probe is moved across a sample while recording a set of far-field diffraction patterns by measuring the intensity of the diffracted waves. The probe is positioned such that each illuminated area has considerable overlap with neighboring regions, providing redundant information that can be used to computationally retrieve {the relative phase of recorded intensity data within the Fraunhofer diffraction plane.} An estimate of the complex image {representing the {refractive index and thickness} of the object} can be obtained from the ptychographic measurements by solving a phase-retrieval optimization problem.
A variety of iterative algorithms have been proposed to solve this problem, the main concepts including batch improvement~\cite{Marchesini.etal2016, Zhai.etal2023, Zhai.etal2021} and stochastic or preconditioned gradient approaches~\cite{Rodenburg.Faulkner2004, Maiden.Rodenburg2009, Maiden.etal2017, Candes.etal2015, Xu.etal2018}.
Although these methods have demonstrated satisfactory performance, they suffer from high computational cost due to their iterative nature.

\emph{Deep learning (DL)} has attracted attention for ptychography due to its potential to reduce the computational cost of ptychographic image reconstruction~\cite{Guan.etal2019, Cherukara.etal2020}. 
Existing techniques depend on \emph{convolutional neural network (CNN)} architectures that directly map measurements to ground truth image patches. Despite being faster than iterative alternatives, CNN-based methods have yet to deliver results comparable with those of iterative methods. This is presumably  because exiting CNNs process individual ptychographic measurements in isolation, thereby preventing the exploitation of the \emph{ptychographic measurement model}, such as the redundant information from overlapping illuminated regions. 
On the other hand, \emph{deep model-based architectures (DMBA)} have shown improved performance over generic CNNs by exploiting the measurement model of imaging problems~\cite{Venkatakrishnan.etal2013, Romano.etal2017, Ongie.etal2020, Monga.etal2021}. A widely-used example of DMBA is the \emph{deep {unrolling} network (DU)} that interprets iterative algorithms as a neural network by stacking iterations into layers and then training it end-to-end. Although DU has shown promising results in many imaging problems, to the best of our knowledge, its potential in the context of ptychographic image reconstruction remains unexplored.

In this paper, we bridge this gap by proposing a novel \emph{{\bf d}eep unrolling network for {\bf ptycho}graphic image reconstruction {based on {\bf v}ision transformer (PtychoDV)}} that leverages the measurement model to improve DL performance while maintaining low computational cost. Our key contributions in this work are summarized as follows:

\begin{itemize}

    \item PtychoDV consists of a \emph{vision transformer (ViT)}~\cite{Dosovitskiy.etal2021} followed by a DU network. ViT employs self-attention mechanisms that learn the interdependencies between measurements and then reconstructs the entire set of data, providing an initial image for DU. This is essential due to the non-convex and nonlinear nature of ptychography{, which} makes {it {nontrivial} to} direct estimation of an initial image from raw data. DU then refines the initial image by alternating between imposing CNN priors and applying the update rule of Wirtinger flow~\cite{Candes.etal2015} based on the measurement model.
  
    \item We tested PtychoDV on simulated data, demonstrating that it (a) achieved state-of-the-art performance compared with DL baselines, (b) obtained competitive results compared with iterative approaches, with substantially reduced computational cost, and (c) has potential for the sparse sampling setup and providing a suitable initialization for iterative methods, even when the probe in testing differs from that in training.
  
\end{itemize}

\begin{figure*}[!]
  \centering
  \includegraphics[width=.865\textwidth]{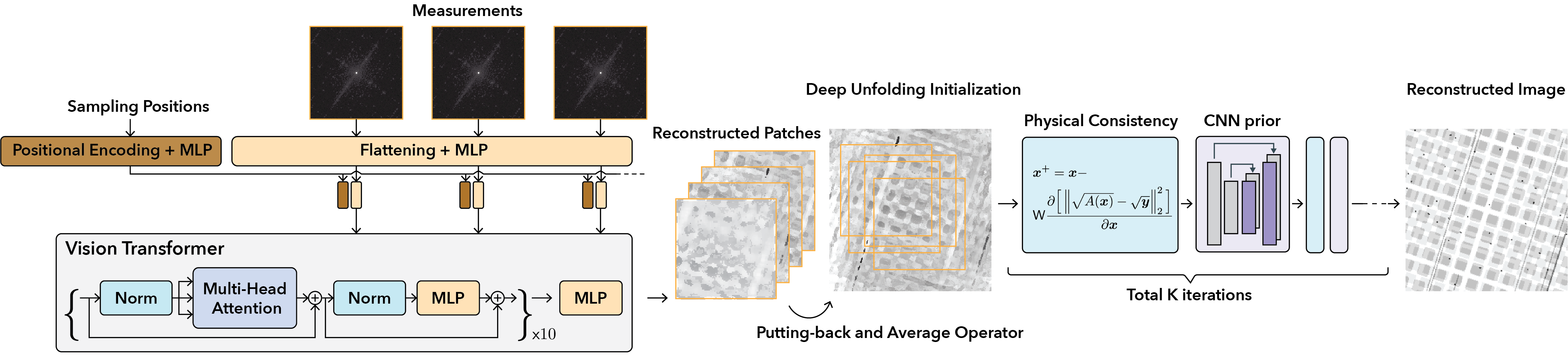}
  \caption{An illustration of the pipeline of PtychoDV that consists of two main components: \emph{(a)} a vision transformer module that reconstructs an initial image from raw measurements by taking into account the interdependencies of the measurements, and \emph{(b)} a DU network that refines the initial image using the measurement forwards and CNN priors. {See \eqref{equ:WF} for the iterative update of the physical consistency module.}}
  \label{fig:method}
\end{figure*}

\section{Related Work}
{In this section, we introduce the notions and related works required to define PtychoDV. We also discuss iterative algorithms and deep learning approaches for ptychographic image reconstruction.}
\subsection{Problem Formulation}
Ptychographic image reconstruction is usually formulated as an inverse problem that recovers an unknown image $\xbm\in\C^n$ from a set of measurements $\{{\ybm_i}\in\R^m\}_i^N$ {characterized} by nonlinear systems
{
\begin{equation}
    \label{equ:fwd}
    {\ybm_i^2\ \sim \mathsf{Pois}(|\Fbm\Pbm\Dbm_i\xbm|^2)}\ .
\end{equation}
}
where $\Pbm\in\C^{m\times m}$ is the complex probe illumination, $\Fbm\in\C^{m\times m}$ represents the Fourier transform, {$\mathsf{Pois(\cdot)}$ denotes a Poisson distribution that models the detector response}, and $|\cdot|$ is an elementwise absolute value operator.
{In this study, we assume the probe is known and only estimates the image.}
In \eqref{equ:fwd}, $\Dbm_i\in{\{0, 1\}}^{m\times n}$ indicates an operator that extracts one patch from $\xbm$, determined by the $i$th probe location during imaging, and $N$ is the total number of probe locations.
{Note that we do not consider subpixel illumination shifts}
For ease of notation in our discussion, we also define $\xbm_i=\Dbm_i\xbm$ as a patch of ground truth corresponding to the $i$th probe location, $\Dbm^{\Tsf}_i\in\C^{n\times m}$ as the adjoint operator of $\Dbm_i$ that transforms a patch into an image by zero-filling the surplus regions.
A common way to solve this inverse problem is to formulate it as an optimization problem
\begin{equation}
    \label{equ:optimization}
    \xbmast = \argmin_\xbm \Big\{\sum_{i=1}^N f_i(\xbm_i)\Big\}\ ,
\end{equation}
where 
\begin{equation}
    \label{equ:fi}
    f_i(\xbm_i) = \frac{1}{2\sigma^2_i}\norm{\ybm_i - |\Fbm\Pbm\xbm_i|}^2
\end{equation}
represents a cost function enforcing data consistency between $\xbm_i$ and $\ybm_i$. This choice of cost function can be derived as an approximation of the maximum likelihood (ML) cost function for a Poisson noise model~\cite{Xu.etal2018}.

\subsection{Iterative Methods}
A variety of numerical iterative algorithms have been proposed for solving \eqref{equ:optimization}~\cite{Marchesini.etal2016, Zhai.etal2023, Zhai.etal2021, Rodenburg.Faulkner2004, Maiden.Rodenburg2009, Maiden.etal2017, Candes.etal2015, Xu.etal2018}.
{Many of these methods concurrently update the image patches $\{\xbmhat_i\}$ and the combine these patches into an estimate image~\cite{Marchesini.etal2016, Zhai.etal2023, Zhai.etal2021}, aiming to overcome the computational challenges posed by the substantial volume of data.
For example, SHARP~\cite{Marchesini.etal2016} relies on alternating projections between constraints in the Fourier domain and image domain. \emph{projected multi-agent consensus equilibrium
(PMACE)}~\cite{Zhai.etal2023, Zhai.etal2021}} solves {ptychography} problem \eqref{equ:optimization} by finding an equilibrium point $\xbm^*$ that satisfies the equation $[F_1(\xbm_1), ..., F_N(\xbm_N)]^T = [\xbmbar_1, ..., \xbmbar_N]^T$, where
\begin{equation}
    F_i(\xbm_i) = \argmin_\vbm \Big\{f_i(\vbm) + \frac{1}{2\sigma^2}\norm{\Fbm\Pbm\vbm - \Fbm\Pbm\xbm_i} \Big\}
\end{equation} 
is derived as a proximal map for $f_i(\xbm_i)$, and
\begin{equation}
    \label{equ:pmace-consensus}
    \xbmbar_i = \Dbm_i\bm{\Lambda}^{-1}\sum_{i=1}^{N}\Dbm^{\Tsf}_i|\Pbm|^\kappa\xbm_i\ 
\end{equation}
appropriately averages the estimated patches associated with the same scan locations. In \eqref{equ:pmace-consensus}, $\bm{\Lambda}=\sum_{i=1}^{N}\Dbm^{\Tsf}_i|\Pbm|^\kappa$, and $\kappa$ denotes a probe exponent parameter. 
Another class of algorithms use stochastic or preconditioned gradient methods to directly refine the estimated image~\cite{Rodenburg.Faulkner2004, Maiden.Rodenburg2009, Maiden.etal2017, Candes.etal2015, Xu.etal2018}. 
For instance, \emph{Wirtinger flow (WF)}~\cite{Candes.etal2015} and \emph{accelerated WF (AWF)}~\cite{Xu.etal2018} use gradient descent to minimize the non-differentiable objective in \eqref{equ:optimization} by defining a generalized gradient based on the notion of Wirtinger derivatives {(see also Sec. VI in ~\cite{Candes.etal2015})}.
While these methods can provide satisfactory performance, they suffer from high computational cost due to the iterative refinement nature.

\subsection{Deep Learning Approaches}
Deep learning has gained popularity in the broader context of imaging inverse problems due to its excellent performance (see recent reviews in~\cite{Ongie.etal2020, mccann2017convolutional, Lucas.etal2018}). A widely-used DL approach is to train a CNN to learn a mapping from the measurements to the desired reconstruction~\cite{Jin.etal2017, Zhu.etal2018}.
Several DL methods based on CNNs have been proposed for ptychographic image reconstruction~\cite{Guan.etal2019,Barutcu.etal2022,Cherukara.etal2020,Guzzi.etal2021}. PtychoNet~\cite{Guan.etal2019} and PtychoNN~\cite{Cherukara.etal2020} involve training an end-to-end DL model by sequentially mapping measurements $\ybm_i$ to corresponding ground truth $\xbm_i$. In testing, one can {derive reconstructed images using the raw measurements as inputs to the pre-trained model.} These methods can achieve fast reconstruction, but at the expense of performance. 
We posit that this is due to CNNs processing individual ptychographic measurements, which fundamentally prevents them from exploiting information from the measurement model, such as the redundancy from the overlapping measured diffraction patterns.
In this study, we propose to tackle these issues by leveraging two {recent} approaches: vision transformer (ViT) and deep model-based architecture (DMBAs), detailed discussions of which follow.

ViTs represents a {significant} shift in computer vision, moving from convolutional architectures to a transformer-based approach (see e.g. recent reviews~\cite{liu2023survey, khan2022transformers}). The central concept behind ViT is treating image patches as data sequences, and then employing \emph{self-attention mechanisms} to compute attention scores among all patch pairs, gauging their reciprocal influence. This approach allows each patch to consider all others in its context, efficiently capturing long-range dependencies and complex interrelationships, irrespective of spatial distance. Recent studies have applied ViT in many imaging inverse problems (see Sec. 3.6 in~\cite{khan2022transformers}). 
In ptychography, it is straightforward to apply ViT by considering measurements as a sequence so that their interdependencies can be learned. Despite that, our empirical results in Table~\ref{tab:exp-noisy} and~\ref{tab:exp-noiseless} show that, while ViT can perform better than CNNs, the performance of ViT is inferior to that of iterative approaches.
{A recent abstract investigated the use of transformers for ptychography~\cite{Kang.etal2021}. Nonetheless, our work distinguishes itself from~\cite{Kang.etal2021} in two key aspects: \emph{(a)} our analysis of the algorithm and numerical validation is more extensive, and \emph{(b)} we improve ViT by integrating DU into our proposed pipeline.}

DMBAs represent a family of DL algorithms that systematically connect measurement models and deep neural networks for solving imaging inverse problems (see also reviews in~\cite{Ongie.etal2020, Monga.etal2021}). 
Examples of DMBAs include \emph{plug-and-play (PnP)}~\cite{Venkatakrishnan.etal2013, Kamilov.etal2023}, \emph{regularization by denoiser (RED)}~\cite{Romano.etal2017}, \emph{deep {unrolling} (DU)}~\cite{Schlemper.etal2018, Hammernik.etal2018, Aggarwal.etal2019, Hu.etal2022, Adler.Oktem2018, Wu.etal2021a, Liu.etal2021a,  Kazemi.etal2022}, and \emph{deep equilibrium models (DEQ)}~\cite{Gilton.etal2021, Liu.etal2022, Gan.etal2022a}.
{PnP and RED represent classes of iterative algorithms that leverage pre-trained denoisers as imaging priors. A recent study has extended this idea to ptychographic image reconstruction~\cite{welker2022deep}. However, its iterative nature inherently results in a high computational cost.}
DU has recently gained significant popularity due to its excellent performance and low computational cost. The key idea of DU is to \emph{(a)} implement a finite number of iterations of an image reconstruction algorithm as layers of a network, \emph{(b)} represent the regularization within the iterative algorithm as a trainable CNN, and \emph{(c)} train the resulting network end-to-end. Many recent studies have shown the potential of DU in various imaging inverse problems, including compressed sensing MRI~\cite{Schlemper.etal2018, Hammernik.etal2018, Aggarwal.etal2019, Hu.etal2022}, sparse view CT~\cite{Adler.Oktem2018, Wu.etal2021a, Liu.etal2021a}, and phase retrieval~\cite{Kazemi.etal2022}. 
{A recent study has explored DU in the context of ptychography~\cite{saha2022lwgnet}, but it lacks a trainable network for providing dedicated initial images.}
Different DU architectures can be obtained by using different iterative algorithms.
As will be discussed in the next section, our main contribution is to propose a deep {unrolling} network based on the WF algorithm to significantly improve the deep learning method performance in ptychographic image reconstruction.

\section{Proposed Method: PtychoDV}
As illustrated in Figure~\ref{fig:method}, PtychoDV consists of two neural networks: (a) a vision transformer $\mathsf{g}_\thetabm$ that estimates initial results from the raw measurements, and (b) a DU network that iteratively refines the initial results. We rely on supervised learning to jointly 
optimize these two neural networks.

\subsection{Vision Transformer}

The vision transformer $\mathsf{g}_\thetabm$ in PtychoDV takes as input a set of raw measurements ${\ybm_i}$ and reconstructs image patches ${\hat{\xbm}_i} = \mathsf{g}_\thetabm({\ybm_i})$. Specifically, the raw measurements are transformed into measurement latent vectors in parallel by a multi-layer perceptron (MLP). The {Cartesian} coordinates of the corresponding sampling position $\cbm$ are mapped to coordinate latent vectors with the same dimension as the measurement latent vectors using Fourier positional encoding~\cite{Mildenhall.etal2020} followed by a MLP
\begin{equation}
    \label{equ:pos_encoding}
    \mathsf{MLP}\big(\sin(2^0\pi\cbm), \cos(2^0\pi\cbm) ... \sin(\underbrace{2^{L_f}\pi}_{k_{\sin}}\cbm), \cos(\underbrace{2^{L_f}\pi}_{k_{\cos}}\cbm)\big)\ ,
\end{equation}
{where $\sin(\cdot)$ and $\cos(\cdot)$ are element-wise operators.}
The measurement feature vectors and the coordinate feature vectors are concatenated and then iteratively processed by attention modules, which consist of layer normalization, multi-head self-attention (MHSA) modules, and MLPs. 
The output feature vectors from the last attention module are transformed into reconstructed patches with the same dimensions as the raw measurements using a output MLP. Further technical details of ViT can be found in~\cite{Dosovitskiy.etal2021}. {The key differences compared to the original VIT~\cite{Dosovitskiy.etal2021} include a different positional embedding derived from the sampling position of ptychography and a modified output layer that transforms the final feature maps of the transformer into patches with dimensions matching those of the raw measurements.}

{The main innovation behind the use of ViT in $\mathsf{g}_\thetabm$ is considering the measurements related to the same ground truth as a sequence. The motivation behind this is to allow the model to capture long-range dependencies and complex relationships among the measurement patches, especially those that overlap, reflecting the imaging nature of ptychography. On the other hand, existing DL methods, such as PtychoNet~\cite{Guan.etal2019}, reconstruct the measurements in parallel, without taking into account dependencies among measurements.}

We then convert the reconstructed patches into an image $\xbmhat$ by the following steps: \emph{(a)} we initialize $\xbmhat$ as an all-zero image and create counters for each pixel location; \emph{(b)} we add each reconstructed patch to the corresponding sampling region in $\xbmhat$ and increase the counters in that area by one; and \emph{(c)} {We perform element-wise division of $\xbmhat$ by the counter at all locations where the counter has non-zero value}.

\subsection{Deep Unrolling Network}
The DU network in PtychoDV is {obtained by interrupting the iteration of} the proximal gradient PnP framework~\cite{Kamilov.etal2023} {which} consists of $K$ iterations of gradient descent each followed by neural network refinement
\begin{equation}
  \label{equ:du}
  \xbmhat^{k+1} = \mathsf{h}_\varphibm\big(\xbmhat^{k} - \mathsf{WF}(\xbmhat^k)\big)\ \forall k=0,\ ...\ ,K-1\ ,
\end{equation}
where $\mathsf{h}_\varphibm$ denotes a CNN with trainable parameter $\varphibm\in\R^m$, $\xbmhat^{k+1}$ is the output of the $k$th layer of DU, and $\xbmhat^0=\xbmhat$. Here, $\mathsf{WF}(\cdot)$ represents a Wirtinger flow gradient update of the objective in \eqref{equ:optimization}
\begin{equation}
  \label{equ:WF}
  \mathsf{WF}(\xbmhat^k)= \gamma \sum_i^N \Dbm^{\Tsf}_i\Pbm^\Hsf\Fbm^\Hsf \left(\mbm_i({\xbmhat^k}) - \ybm_i\frac{{\mbm_i(\xbmhat^k)}}{|{\mbm_i(\xbmhat^k)}|}\right)\ ,
\end{equation}
where
{
\begin{equation}
    \mbm_i(\xbmhat^k) = \Fbm\Pbm\Dbm_i\xbmhat^k \;,
\end{equation}
$(\cdot)^\Hsf$ is the conjugate transpose,} and $\gamma$ represents a step size of $\max(\sum_{i=1}^{N}\Dbm^{\Tsf}_i|\Pbm|^2)$. The WF gradient descent allows DU to exploit the information from the physical model of ptychography by fitting the intermediate estimation to the raw measured data. {$\mathsf{h}_\varphibm$ further refines the estimation by imposing a prior information learned from the external dataset.}

\subsection{Loss Function}
We trained $\mathsf{g}_\thetabm$ and $\mathsf{h}_\varphibm$ jointly in an end-to-end manner by minimizing the loss function
\begin{equation}
  \label{equ:loss}
  \ell_\mathsf{loss} = \ell_\mathsf{image}(\varphibm) + \lambda\ \ell_\mathsf{patch}(\thetabm)\ ,
\end{equation}
where $\lambda$ is a trade-off parameter. The purpose of \eqref{equ:loss} is to promote high-quality reconstruction in both the image-wise and patch-wise manners. Specifically, $\ell_\mathsf{image}$ is formulated to penalize the difference between the final estimation of DU and the corresponding ground truth
\begin{equation}
  \label{equ:loss-image}
  \ell_\mathsf{image} = \norm{\xbmhat^K - \xbm}^2\ ,
\end{equation}
and $\ell_\mathsf{patch}$ seeks to minimize the discrepancy between estimated patches of ViT and the corresponding ground truth patches
\begin{equation}
  \label{equ:loss-patch}
  \ell_\mathsf{patch} = \sum_{i=1}^N\norm{\xbmhat_i - \xbm_i}^2\ .
\end{equation}

\begin{figure}
  \centering
  \includegraphics[width=.385\textwidth]{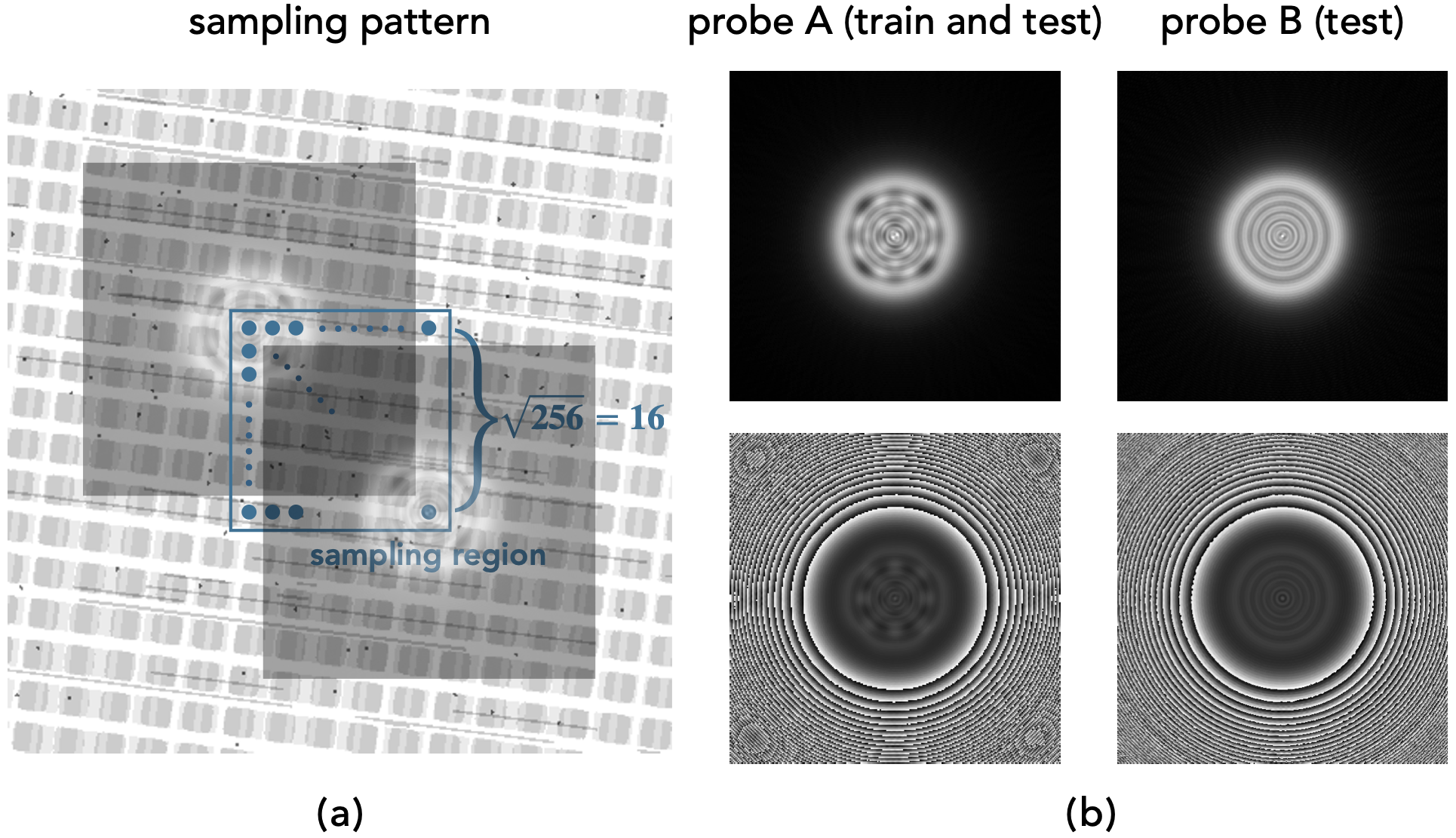}
  \caption{Illustrations of magnitude of ground truth {image}, two simulated ground truth probes (images in the top row are magnitude, bottom row phase), and sampling pattern of {$256$:$5$}. \emph{Probe A} was used to synthesize measurement for training and testing, while \emph{probe B} was exclusively for testing the pre-trained models.}
  \label{fig:groundtruths}
\end{figure}

\begin{table*}[]
  \scriptsize
  \caption{Quantitative evaluation of several methods with format of $A\pm B(c)$ on testing noisy measurements, where $A$, $B$ and $c$ denote mean of normalized root mean-square-error (NRMSE), standard deviation of NRMSE, and testing time (seconds per image), respectively. The results with {\bf \color{red}the best} and \underline{\color{blue}second best} mean NRMSE are highlighted. This table shows that PtychoDV can outperform existing DL baseline methods. This table also demonstrates that PtychoDV can gain competitive performance compared against state-of-the-art iterative algorithm, while maintaining significantly lower computational cost.}
  \label{tab:exp-noisy}
  \centering
  \renewcommand\arraystretch{1.1}
  \setlength{\tabcolsep}{0.75pt}
  \begin{threeparttable}
    \begin{tabular}{llllll}
      \toprule
      Sampling pattern & $256$:$5$   & $121$:$8$  & $64$:$11$  & $25$:$19$  & $16$:$27$  \\
      \cmidrule{2-6}
      PtychoNet~\cite{Guan.etal2019}  & 0.483 ± 0.56 (0.175)  & 0.483 ± 0.56 (0.075) & 0.483 ± 0.56 (0.042) & 0.483 ± 0.56 (0.017) & 0.484 ± 0.56 (0.012) \\
      Unet~\cite{Ronneberger.etal2015}       & 0.465 ± 0.55 (0.366)  & 0.465 ± 0.55 (0.165) & 0.465 ± 0.55 (0.084) & 0.466 ± 0.55 (0.034) & 0.467 ± 0.55 (0.022) \\
      ViT~\cite{Dosovitskiy.etal2021}        & 0.441 ± 0.59 (0.062)  & 0.442 ± 0.59 (0.030) & 0.443 ± 0.59 (0.020) & 0.447 ± 0.59 (0.009) & 0.450 ± 0.59 (0.009) \\
      AWF~\cite{Xu.etal2018}        & 0.047 ± 0.15 (109.60) & 0.054 ± 0.20 (51.29) & 0.071 ± 0.21 (26.12) & 0.118 ± 0.34 (10.70) & 0.201 ± 0.63 (7.15)  \\
      PMACE~\cite{Zhai.etal2021}      & \textbf{\color{red} 0.035 ± 0.18 (138.08)} & \textbf{\color{red} 0.044 ± 0.13 (66.43)} & \textbf{\color{red} 0.065 ± 0.19 (34.94)} & 0.119 ± 0.33 (13.82)  & 0.184 ± 0.52 (8.78) \\
      \cmidrule{2-6}
      ViT+Unet & 0.219 ± 0.65 (0.059)  & 0.241 ± 0.68 (0.025) & 0.254 ± 0.64 (0.017) & 0.284 ± 0.65 (0.010) & 0.307 ± 0.67 (0.011) \\
      ViT+GD  & 0.259 ± 0.38 (0.177)  & 0.261 ± 0.38 (0.081) & 0.281 ± 0.41 (0.045) & 0.897 ± 0.30 (0.021) & 0.929 ± 0.32 (0.015) \\
      ViT+1DU  & 0.128 ± 0.54 (0.118)  & 0.139 ± 0.56 (0.051) & 0.164 ± 0.56 (0.031) & 0.218 ± 0.64 (0.016) & 0.245 ± 0.63 (0.015) \\
      {Initializer+DU} & {0.069 ± 0.25 (0.362)}  & {0.076 ± 0.28 (0.183)} & {0.093 ± 0.34 (0.122)} & {0.137 ± 0.41 (0.085)} & {0.173 ± 0.49 (0.077)} \\
      PtychoNet+DU & 0.046 ± 0.19 (0.617)    & 0.053 ± 0.22 (0.344)   & \underline{\color{blue}0.066 ± 0.30 (0.265)}   & \underline{\color{blue}0.104 ± 0.37 (0.245)}   & \underline{\color{blue}0.139 ± 0.46 (0.239)} \\
      \cmidrule{2-6}
      PtychoDV       & \underline{\color{blue}0.043 ± 0.19 (0.212)} & \underline{\color{blue}0.050 ± 0.23 (0.109)} & \textbf{\color{red} 0.065 ± 0.32 (0.074)} & \textbf{\color{red} 0.098 ± 0.36 (0.049)} & \textbf{\color{red} 0.127 ± 0.45 (0.044)} \\
      \bottomrule       
    \end{tabular}
  \end{threeparttable}
\end{table*}

\begin{figure*}
  \centering
  \includegraphics[width=.775\textwidth]{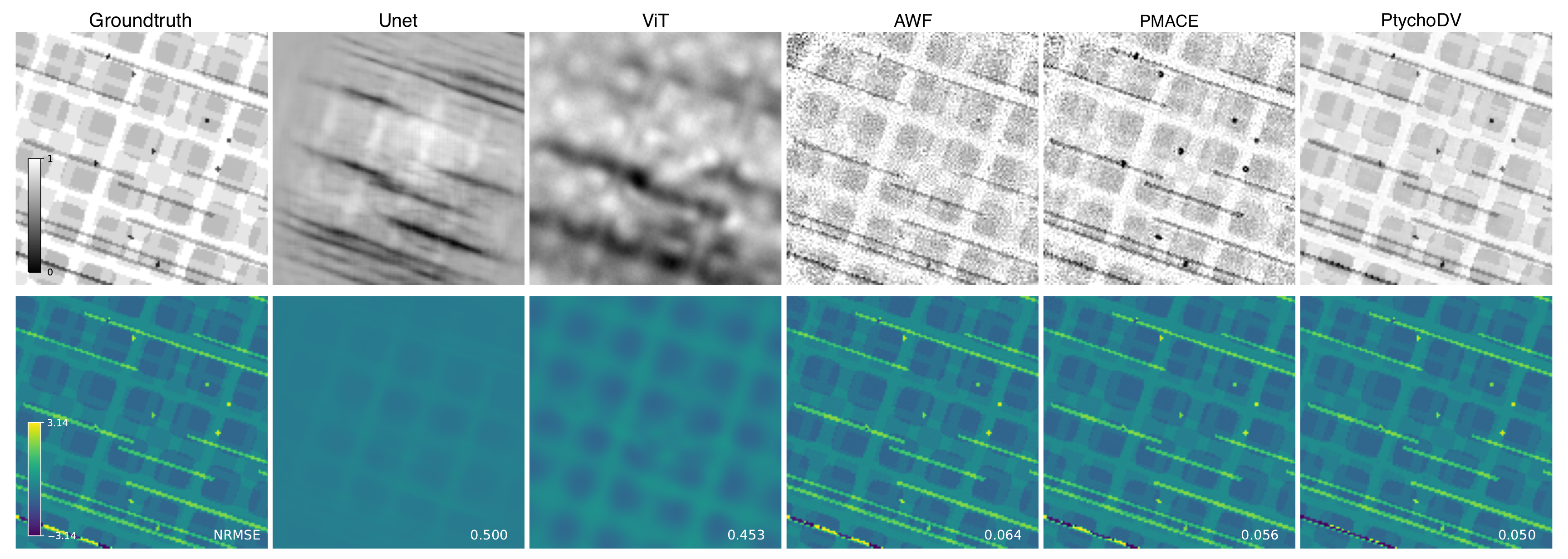}
  \caption{Visual results of PtychoDV and other baseline methods on noisy testing data with sampling pattern of $64$:$11$. The magnitude and the phase of the reconstructed images are shown in the top and the bottom row, respectively. NRMSE values are included in the right bottom of each image. This figure highlights superior performance of PtychoDV on sparse sampling pattern. Note that PtychoDV can reconstruct images that are consistent with ground truth, {whereas the results from the other baseline exhibit noise and blurry artifacts.}}
  \label{fig:baseline-noisy}
\end{figure*}

\section{Numerical Validation}
{This section presents {the} setup and results of our numerical validation on PtychoDV. {We discuss our dataset, the implementation of PtychoDV, our comparison method, and our evaluation metrics.}}
\subsection{{Experimental} Setup} 
\label{sec:exp-setup}
\subsubsection{Dataset} 
{We {simulated a} dataset consisting of ground truth complex-valued images (\emph{i.e.,} $\xbm$ in \eqref{equ:fwd}) and ground truth complex-valued probes (\emph{i.e.,} $\Pbm$ in \eqref{equ:fwd}).
Ground truth images were {$400 \times 400$} pixels and had assigned density and thickness to model a multi-layer Copper-Tungsten composite material. 
Simulated probes were $256 \times 256$ pixels {with a photon energy of 8.8 keV}.} We simulated 60,000, 100, and 100 ground truth samples for training, validation, and testing, respectively.
We simulated two types of probes, which we shall refer to as \emph{probe A} and \emph{probe B}. We used \emph{probe A} to generate datasets for training and testing, while \emph{probe B} was used only for testing, in order to evaluate the generalization of the pre-trained model on measurements simulated using an unseen probe.
\emph{Probe B} was assumed to be unknown in this experiment, while \emph{probe A} was known.
Different sampling patterns (i.e., $\Dbm_i$ in Equation \eqref{equ:fwd}) were simulated, denoted as {$N$:$L$, where the probe locations form an $\sqrt{N}\times \sqrt{N}$ grid with {grid spacing} equal to $L$ {pixels}. {We experimented with $N$:$L$ values of} $256$:$5$, $121$:$8$, $64$:$11$, $25$:$19$, and $16$:$27$.} The smaller the value of $N$, the sparser the sampling pattern.
The training dataset involves different sampling patterns.
Figure~\ref{fig:groundtruths} illustrates a sample of ground truth images, $256$:$5$ sampling patterns, and the simulated ground truth probes.
We followed~\cite{Zhai.etal2021} to use $r_p$, the peak photon rate, to scale the mean of a Poisson distribution to obtain noisy simulated measurements
{
\begin{equation}
  \hat{\ybm}_i^2 \sim \mathsf{Pois}\left(\frac{|\Fbm\Pbm\xbm_i|^2}{\mathsf{max}(|\Fbm\Pbm\xbm_i|^2)}\times r_p\right)\ .
\end{equation}
}
As $r_p$ increases, the signal-to-noise ratio also increases. {Assuming a photon detector with 14-bit dynamic range, we} take $r_p=10^5$ for our simulated noisy diffraction patterns. Figure~\ref{fig:groundtruths} illustrates a sample of ground truth images, two stimulated ground truth probes, and sampling pattern of $256$:$5$.

\subsubsection{Implementation}
We experimented with several values of $\lambda$ in \eqref{equ:loss}. The best empirical results were obtained when $\lambda=1$. We set the number of DU iteration of PtychoDV to $K=3$, which is the maximum number achievable under the memory constraints of our workstation.
{We set $L_f$ in \eqref{equ:pos_encoding} to 10.}
We used the Adam~\cite{Kingma.Ba2017} optimizer with learning rate {$10^{-5}$} and mini-batch size $1$, training for $30$ epochs. We performed all experiments on a host equipped with an AMD Ryzen Threadripper 3960X Processor and an NVIDIA GeForce RTX 3090 GPU. {The training time of PtychoDV on this host was around 120 hours.} Our PtychoDV implementation is publicly available\footnote{\url{https://github.com/wjgancn/PtychoDV}}.

\begin{figure*}
  \centering
  \includegraphics[width=.675\textwidth]{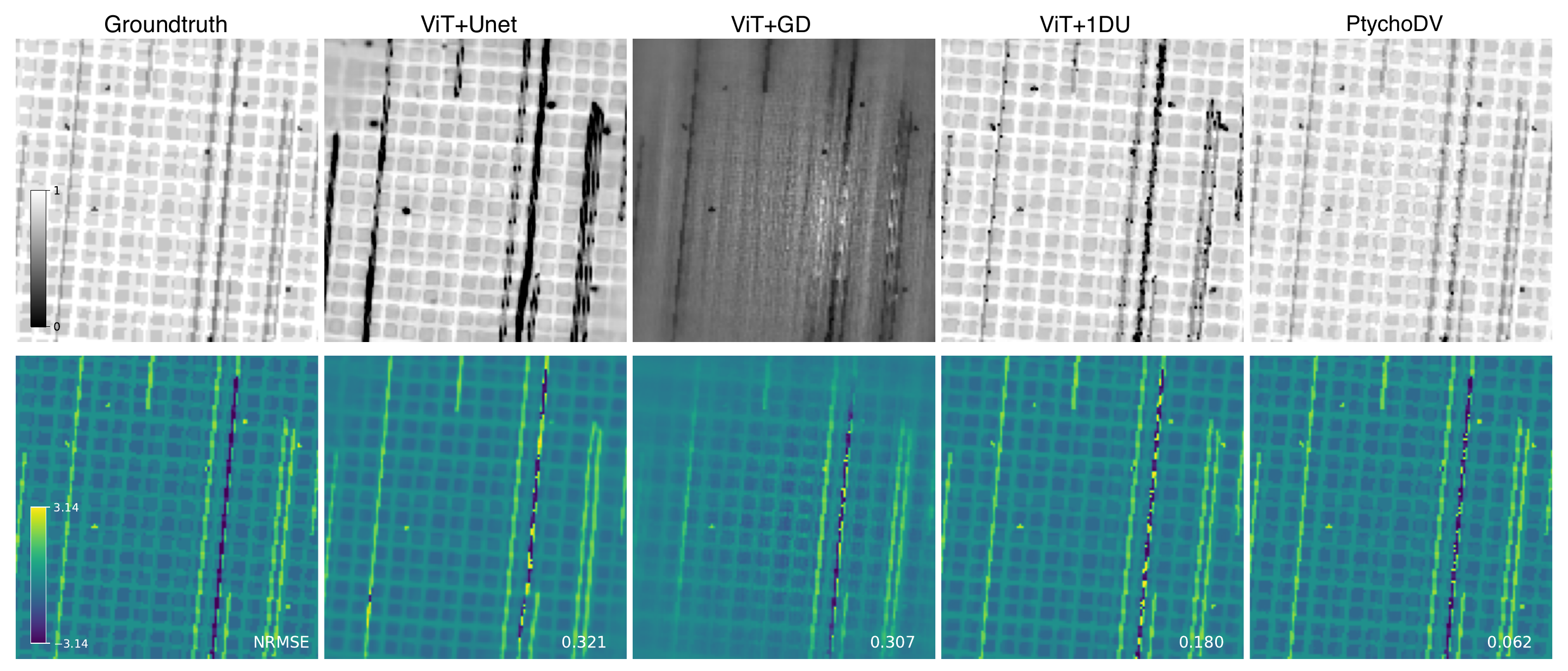}
  \caption{Visual results of PtychoDV and its variants on noisy testing data with sampling pattern of $64$:$11$. The magnitude and the phase of the reconstructed images are shown in the top and the bottom row, respectively. NRMSE values of each method is labeled in the right bottom of each image. This figure shows that PtychoDV can gain superior performance over its ablated methods.}
  \label{fig:ablation-noisy}
\end{figure*}

\subsubsection{Evaluation}
We followed~\cite{Zhai.etal2021} in using normalized root-mean-square error (NRMSE) to evaluate the quality of reconstructed images. Because that the measured data is not sensitive to a constant phase shift in the full transmittance image, we have taken into account this phase shift while calculating the NRMSE between the reconstructed complex image $\xbmhat$ and the ground truth image $\xbm$. Specifically, the NRMSE is calculated elementwise as follows:
\begin{equation}
  \mathsf{NRMSE}(\hat{x}, x) = \frac{|{\hat{x} - e^{i\theta}x}|}{|{x}|} \ ,
\end{equation}
where $\theta\in[0, 2\pi)$ is chosen to minimize the numerator.

\subsubsection{Comparison}
We compared PtychoDV with several baseline approaches, including PtychoNet~\cite{Guan.etal2019}, Unet, ViT, PMACE~\cite{Zhai.etal2021} and AWF~\cite{Xu.etal2018}. PtychoNet is a DL method that uses a CNN to map individual measurements directly to the corresponding ground truth image patch. 
{We implemented PtychoNet, Unet, and ViT. For PMACE and AWF, we used the official implementations from the PMACE repository\footnote{\url{https://github.com/cabouman/ptycho_pmace}}.}
We set the total number of iterations of PMACE and AWF to 100. We followed~\cite{Zhai.etal2021} to estimate the initial images for PMACE and AWF. Unet and ViT is similar to PtychoNet, but having more complex neural network architectures.

In order to determine the impact of different elements in our configuration, we conducted a component analysis with various versions of PtychoDV, termed as \emph{ViT+Unet}, \emph{ViT+GD}, \emph{ViT+1DU}, {\emph{Initializer+DU}} and \emph{PtychoNet+DU}. \emph{ViT+Unet} replaces the DU with Unet, thereby removing the integration of the measurement models in the resulting network architecture. \emph{ViT+GD} excludes the CNN priors in DU, whereas \emph{ViT+1DU} reduces the number of DU iterations to one. \emph{PtychoNet+DU} substitutes ViT with PtychoNet as the CNN used for computing the initial images. {\emph{Initializer+DU} substitutes ViT with a handcrafted initialization approach (refer to equation (24) in \cite{Zhai.etal2021}). The trainable components of \emph{Initializer+DU} constitute a pure deep unrolling architecture.
} 

{In addition, we tested the use of the PtychoDV reconstructions as initialization for PMACE}. We conducted experiments on both \emph{probe A} and \emph{probe B}. The resulting methods are as follows: \emph{(a) PtychoDV-A} tests PtychoDV on testing data stimulated using \emph{probe A}; \emph{(b) PMACE-A} tests PMACE on testing data stimulated using \emph{probe A}; \emph{(c) PMACE-A-10} is a variant of \emph{PMACE-A} with total number of iterations being 10; \emph{(d) PMACE-A-10 w/ PtychoDV} is similar to \emph{PMACE-A-10} but use PtychoDV to estimate the initial image; \emph{(e) PtychoDV-B} tests PtychoDV on testing data stimulated using \emph{probe B}; \emph{(f) PMACE-B} tests PMACE on testing data stimulated using \emph{probe B}; \emph{(g) PMACE-B w/ PtychoDV} is similar to \emph{PMACE-B} but use PtychoDV to estimate the initial image. Since \emph{probe B} was assumed to be unknown, PMACE in (f) and (g) was implemented to jointly estimate the image and the probe. {We used probe A as the initial probe when jointly estimating probe B.}

\begin{table*}[]
  \scriptsize
  \caption{Quantitative evaluation of several methods with format of $A\pm B(c)$ on testing noisy measurements, where $A$, $B$ and $c$ denote mean of normalized root mean-square-error (NRMSE), standard deviation of NRMSE, and testing time (seconds per image), respectively. The results with {\bf \color{red}the best} and \underline{\color{blue}second best} mean NRMSE over the same testing data are highlighted. This table highlights that PtychoDV initialization could significantly reduce number of iteration of PMACE, thus reducing the computational cost, without sacrificing the performance. This table also shows that PtychoDV could also provide good initialization for better imaging quality of PMACE even when the testing probe differs to that used for training.}
  \label{tab:exp-initializer}
  \centering
  \renewcommand\arraystretch{1.1}
  \setlength{\tabcolsep}{0.75pt}
  \begin{threeparttable}
    \begin{tabular}{llllll}
      \toprule
      Sampling pattern & $256$:$5$   & $121$:$8$  & $64$:$11$  & $25$:$19$  & $16$:$27$  \\
      \cmidrule{2-6}
      PtychoDV-A       & 0.044 ± 0.17 (0.212) & 0.053 ± 0.23 (0.109) & 0.066 ± 0.27 (0.074) & \underline{\color{blue}0.102 ± 0.40 (0.049)} & \underline{\color{blue}0.135 ± 0.49 (0.044)} \\
      PMACE-A     & {\bf \color{red}0.035 ± 0.17 (138.08)} & \underline{\color{blue}0.045 ± 0.18 (66.43)} & \underline{\color{blue}0.065 ± 0.20 (34.94)} & 0.118 ± 0.36 (13.82)  & 0.193 ± 0.58 (8.78)  \\
       PMACE-A-10 & 0.246 ± 0.61 (16.22)   & 0.251 ± 0.61 (8.16)   & 0.264 ± 0.62 (4.62)   & 0.297 ± 0.65 (2.29)   & 0.351 ± 0.74 (1.72)   \\
       \makecell[l]{PMACE-A-10 \\ \ w/ PtychoDV}  & \underline{\color{blue}0.039 ± 0.09 (16.43)}   & {\bf \color{red}0.042 ± 0.08 (8.43)}   & {\bf \color{red}0.048 ± 0.08 (5.10)}   & {\bf \color{red}0.066 ± 0.07 (2.34)}   & {\bf \color{red}0.084 ± 0.15 (1.79)}   \\
       \cmidrule{2-6}
       PtychoDV-B            & \underline{\color{blue}0.151 ± 0.57 (0.212)} & \underline{\color{blue}0.157 ± 0.55 (0.109)} & \underline{\color{blue}0.176 ± 0.65 (0.074)} & \underline{\color{blue}0.209 ± 0.59 (0.049)} & \underline{\color{blue}0.245 ± 0.75 (0.044)} \\
       PMACE-B  & 0.332 ± 0.81 (288.91)   & 0.329 ± 0.84 (137.25)   & 0.340 ± 0.82 (73.12)   & 0.412 ± 0.71 (28.51)   & 0.438 ± 0.64 (17.84)   \\
       \makecell[l]{PMACE-B w/ \\ \ PtychoDV}   & {\bf \color{red}0.081 ± 0.51 (291.54)}   & {\bf \color{red}0.089 ± 0.34 (139.49)}   & {\bf \color{red}0.096 ± 0.53 (75.76)}   & {\bf \color{red}0.139 ± 0.42 (28.93)}   & {\bf \color{red}0.166 ± 0.64 (18.13)}   \\
      \bottomrule
      \end{tabular}
  \end{threeparttable} 
\end{table*}

\subsection{Results}
\label{sec:exp-result}
Table~\ref{tab:exp-noisy} provides a quantitative evaluation and testing time for PtychoDV, baseline approaches, and {methods with different components} during testing of noisy cases with all sampling patterns. As displayed in Table~\ref{tab:exp-noisy}, ViT can achieve lower average NRMSE values than Unet and PtychoNet, both of which are CNN-based, but it still performs suboptimally when compared to iterative methods. When comparing PtychoDV with ViT+Unet and ViT+GD, it is evident from Table~\ref{tab:exp-noisy} that DU and $\mathsf{h}_\varphibm$ are essential components of PtychoDV for achieving superior imaging quality. The results from ViT+1DU indicate the potential for improving PtychoDV by increasing the number of DU iterations. 
{Table~\ref{tab:exp-noisy} shows that, when comparing with \emph{Initializer+DU} and \emph{PtychoNet+DU}, PtychoDV can gain superior performnace, highlighting the importance of using ViT to compute the initial image. While both ViT and DU serve as necessary components within PtychoDV, Table~\ref{tab:exp-noisy} indicates that incorporating DU leads to higher SNR improvements than incorporating ViT. To conlude,} Table~\ref{tab:exp-noisy} demonstrates that PtychoDV can achieve performance that is competitive with, and even superior to (in the sparse sampling pattern), PMACE, the state-of-the-art iterative method.
{Finally,} while PtychoDV is the most time-consuming method among DL baselines, it still has significantly less computational cost than iterative methods across all sampling patterns. 

Figure~\ref{fig:baseline-noisy} provides visual results of PtychoDV and baseline methods on noisy cases with \emph{sparse} $64$:$11$ sampling patterns. 
Figure~\ref{fig:baseline-noisy} shows that end-to-end neural networks, which directly map measurements to ground truth image patches, tend to reconstruct images with blurry details, whereas PtychoDV provides {less noisy} and {sharper} images. This figure also highlights that AWF and PMACE, the two commonly used iterative algorithms, reconstruct noisy images with a higher NRMSE than PtychoDV in the sparse $64$:$11$ sampling pattern. Figure~\ref{fig:ablation-noisy} provides visual results of PtychoDV compared to ablated methods on testing noisy cases with the sampling patterns of $64$:$11$. This figure illustrates that PtychoDV can quantitatively and qualitatively outperform several ablated variants.

{Table~\ref{tab:exp-initializer} provide a quantitative evaluation and testing time for PtychoDV and PMACE on noisy testing data, simulated with different probes and different sampling patterns.} These tables demonstrate that, when tested on a \emph{known} probe A, PMACE initialized by PtychoDV can achieve performance competitive with generic PMACE, but with significantly fewer iterations and lower computational cost. The tables also indicate that, when tested on an \emph{unknown} probe B, PMACE initialized by PtychoDV can achieve superior performance compared to PMACE on the joint estimation of image and probe. 

\section{Discussion and Conclusion}
This paper presents PtychoDV, a new DL method for ptychographic image reconstruction. The key idea behind PtychoDV is a deep {unrolling} architecture that systematically integrates trainable neural network priors and measurement operators of the ptychography. Moreover, we employ a vision transformer to estimate initial images from raw measurements, which allows capturing long-range dependencies in the data effectively.

The major benefits of PtychoDV include its remarkable performance improvements, both quantitatively and qualitatively, compared to existing deep learning methods. Furthermore, PtychoDV achieves competitive performance against existing iterative algorithms, but with a substantially lower computational cost. Moreover, in sparse sampling setup, PtychoDV outperforms iterative methods. The results of PtychoDV show its potential for applications that require real-time reconstruction or fast sampling.

Another {important application} of PtychoDV is to provide a reliable initialization for existing iterative algorithms. This {initialization approach} leads to a reduction in the total number of iterations without sacrificing performance. Even in cases where the probe is unknown, iterative algorithms can still benefit from PtychoDV's initialization, regardless of whether the testing probe differs from the training one.

{A key feature of PtychoDV is its ability to incorporate and exchange information from all measurements patches simultaneously in the reconstruction. This exchange is technically facilitated through the WF update as described in \eqref{equ:WF} and the attention module in ViT. Note that the effectiveness of this exchange is also contingent on the overlap probe ratio. Experimental results show that a higher overlap ratio (\emph{e.g.,} sampling pattern of $256:5$) leads to improved performance, indicating enhanced exchange efficiency. In contrast, existing approaches without such deliberate exchange (\emph{e.g.,} PtychoNet) achieve similar performance across different sampling patterns.}

The experiments in this study were entirely simulation-based, primarily due to the large number of training pairs and high-quality references required for the proposed method. It {is} impractical to source such a dataset from real-world samples. Our future direction includes {testing} PtychoDV on real data and training PtychoDV without high-quality ground truth using self-supervised learning~\cite{Hu.etal2022}.

\section{Acknowledgements}
This research was supported by the Laboratory Directed Research and Development program of Los Alamos National Laboratory under project number 20230771DI. The authors thank John Barber for his assistance in using the LANL wave propagation code WavePro to generate the training data used in this work.

\begin{table*}[]
  \scriptsize
  \caption{Quantitative evaluation of several methods with format of $A\pm B(c)$ on testing \emph{noise-free} measurements, where $A$, $B$ and $c$ denote mean of normalized root mean-square-error (NRMSE), standard deviation of NRMSE, and testing time (seconds per image), respectively. The results with {\bf \color{red}the best} and \underline{\color{blue}second best} mean NRMSE are highlighted. As evidenced by the table, PtychoDV surpasses existing deep learning baseline methods. Moreover, it showcases that PtychoDV can achieve performance comparable to state-of-the-art iterative algorithms while maintaining considerably lower computational cost.}
  \label{tab:exp-noiseless}
  \centering
  \renewcommand\arraystretch{1.1}
  \setlength{\tabcolsep}{0.75pt}
  \begin{threeparttable}
    \begin{tabular}{llllll}
      \toprule
      Sampling pattern & $256$:$5$   & $121$:$8$  & $64$:$11$  & $25$:$19$  & $16$:$27$  \\
      \cmidrule{2-6}
      PtychoNet~\cite{Guan.etal2019} & 0.488 ± 0.55 (0.174) & 0.488 ± 0.55 (0.075) & 0.488 ± 0.55 (0.040) & 0.488 ± 0.55 (0.017) & 0.489 ± 0.55 (0.012) \\
      Unet~\cite{Ronneberger.etal2015}      & 0.455 ± 0.58 (0.362) & 0.455 ± 0.58 (0.164) & 0.455 ± 0.58 (0.084) & 0.456 ± 0.58 (0.034) & 0.456 ± 0.58 (0.022) \\
      ViT~\cite{Dosovitskiy.etal2021}        & 0.410 ± 0.59 (0.061) & 0.411 ± 0.59 (0.030) & 0.413 ± 0.59 (0.018) & 0.415 ± 0.59 (0.011) & 0.416 ± 0.59 (0.008) \\
      AWF~\cite{Xu.etal2018}        & \underline{\color{blue}0.012 ± 0.13 (55.32)} & \underline{\color{blue}0.013 ± 0.22 (26.61)} & 0.023 ± 0.22 (14.22) & 0.048 ± 0.41 (5.77)  & 0.107 ± 0.75 (3.94)  \\
      PMACE~\cite{Zhai.etal2021}      & \textbf{\color{red} 0.005 ± 0.10 (76.79)} & \textbf{\color{red} 0.006 ± 0.06 (36.67)} & \textbf{\color{red} 0.010 ± 0.11 (19.80)} & \textbf{\color{red} 0.022 ± 0.26 (7.72)}  & 0.044 ± 0.48 (5.08)  \\
      \cmidrule{2-6}
       ViT+Unet     & 0.174 ± 0.61 (0.056) & 0.188 ± 0.60 (0.025) & 0.209 ± 0.60 (0.018) & 0.240 ± 0.60 (0.012) & 0.257 ± 0.62 (0.009) \\
       ViT+GD      & 0.761 ± 0.51 (0.174) & 0.772 ± 0.52 (0.080) & 0.767 ± 0.49 (0.046) & 0.897 ± 0.30 (0.021) & 0.929 ± 0.32 (0.015) \\
       ViT+1DU      & 0.081 ± 0.36 (0.117) & 0.094 ± 0.38 (0.049) & 0.116 ± 0.47 (0.030) & 0.157 ± 0.56 (0.016) & 0.178 ± 0.59 (0.013) \\
       {Initializer+DU} & {0.051 ± 0.22 (0.362)} & {0.052 ± 0.27 (0.183)} & {0.063 ± 0.34 (0.122)} & {0.097 ± 0.56 (0.085)} & {0.131 ± 0.71 (0.077)} \\
       PtychoNet+DU & 0.034 ± 0.15 (0.617) & 0.034 ± 0.17 (0.344) & 0.040 ± 0.21 (0.265) & 0.059 ± 0.37 (0.245) & 0.084 ± 0.53 (0.239) \\
      \cmidrule{2-6}
       PtychoDV           & 0.013 ± 0.10 (0.211) & \underline{\color{blue}0.013 ± 0.10 (0.110)} & \underline{\color{blue}0.017 ± 0.14 (0.074)} & \underline{\color{blue}0.028 ± 0.23 (0.049)} & \textbf{\color{red} 0.038 ± 0.28 (0.044)} \\
      \bottomrule       
    \end{tabular}
  \end{threeparttable}
\end{table*}

\begin{table*}[]
  \scriptsize
  \caption{Quantitative evaluation of several methods with format of $A\pm B(c)$ on testing \emph{noise-free} measurements, where $A$, $B$ and $c$ denote mean of normalized root mean-square-error (NRMSE), standard deviation of NRMSE, and testing time (seconds per image), respectively. The results with {\bf \color{red}the best} and \underline{\color{blue}second best} mean NRMSE over the same testing data are highlighted. The information provided in this table demonstrates that by utilizing PtychoDV for initializations, the number of iterations required by PMACE can be significantly reduced, thus lowering computational cost without compromising the performance. It also underlines that even when the probe used for testing is different from the one used during training, PtychoDV can still provide beneficial initialization to enhance PMACE's imaging quality.}
  \label{tab:exp-initializer-noiseless}
  \centering
  \renewcommand\arraystretch{1.1}
  \setlength{\tabcolsep}{0.75pt}
  \begin{threeparttable}
    \begin{tabular}{llllll}
      \toprule
      Sampling pattern & $256$:$5$   & $121$:$8$  & $64$:$11$  & $25$:$19$  & $16$:$27$  \\
      \cmidrule{2-6}
      PtychoDV-A       & 0.014 ± 0.09 (0.211) & 0.014 ± 0.12 (0.110) & 0.018 ± 0.13 (0.074) & 0.030 ± 0.27 (0.049) & \underline{\color{blue}0.042 ± 0.32 (0.044)} \\
      PMACE-A-10 & 0.181 ± 0.65 (16.22) & 0.188 ± 0.66 (8.16)  & 0.210 ± 0.67 (4.62)  & 0.252 ± 0.70 (2.29)  & 0.316 ± 0.79 (1.72)  \\
      PMACE-A      & \underline{\color{blue}0.006 ± 0.10 (76.79)} & \underline{\color{blue}0.008 ± 0.14 (36.67)} & \underline{\color{blue}0.012 ± 0.13 (19.80)} & \underline{\color{blue}0.025 ± 0.30 (7.72)}  & 0.048 ± 0.48 (5.08)  \\
      \makecell[l]{PMACE-A-10 \\ \ w/ PtychoDV}  & {\bf \color{red}0.003 ± 0.02 (16.43)} & {\bf \color{red}0.003 ± 0.02 (8.43)}  & {\bf \color{red}0.004 ± 0.03 (5.10)}  & {\bf \color{red}0.006 ± 0.04 (2.34)}  & {\bf \color{red}0.010 ± 0.06 (1.79)}  \\
      \cmidrule{2-6}
      PtychoDV-B            & \underline{\color{blue}0.109 ± 0.42 (0.212)} & \underline{\color{blue}0.111 ± 0.49 (0.109)} & \underline{\color{blue}0.120 ± 0.49 (0.074)} & \underline{\color{blue}0.147 ± 0.72 (0.049)} & \underline{\color{blue}0.163 ± 0.65 (0.044)} \\
      PMACE-B  & 0.321 ± 0.80 (288.91)   & 0.317 ± 0.82 (137.25)   & 0.326 ± 0.80 (73.12)   & 0.391 ± 0.76 (28.51)   & 0.421 ± 0.69 (17.84)   \\
      \makecell[l]{PMACE-B \\ \ w/ PtychoDV}   & {\bf \color{red}0.017 ± 0.16 (291.54)}   & {\bf \color{red}0.021 ± 0.40 (139.49)}   & {\bf \color{red}0.019 ± 0.19 (75.76)}   & {\bf \color{red}0.045 ± 0.50 (28.93)}   & {\bf \color{red}0.045 ± 0.43 (18.13)}   \\
      \bottomrule
      \end{tabular}
  \end{threeparttable}
\end{table*}

\begin{table}[]
{\caption{
GPU memory usage (GB) for PtychoDV and DL baseline methods during both training and inference.}
\label{tab:gpu-usage}
\scriptsize
\centering
  \renewcommand\arraystretch{1.1}
  \setlength{\tabcolsep}{5pt}
\begin{center}
\begin{tabular}{lllllll}
\toprule
                 & \multicolumn{5}{l}{Inference}    & Training \\
\cmidrule{2-6}
Sampling Pattern & $256$:$5$   & $121$:$8$  & $64$:$11$  & $25$:$19$  & $16$:$27$    \\
\cmidrule{2-7}
PtychoNet~\cite{Guan.etal2019}        & 11.84 & 6.61  & 4.40  & 2.88  & 2.53  & 21.01         \\
Unet~\cite{Ronneberger.etal2015}             & 20.13 & 10.51 & 5.68  & 3.44  & 2.88  & 23.12         \\
ViT~\cite{Dosovitskiy.etal2021}              & 4.09  & 3.71  & 3.55  & 3.45  & 3.41  & 14.57         \\
\cmidrule{2-7}
ViT+Unet         & 5.73  & 4.81  & 4.47  & 4.57  & 4.51  & 15.46         \\
ViT+GD           & 4.81  & 4.08  & 3.77  & 3.59  & 3.51  & 14.67         \\
ViT+1DU          & 5.45  & 4.73  & 4.54  & 4.63  & 4.56  & 15.47         \\
PtychoNet+DU     & 11.94 & 6.71  & 4.49  & 2.98  & 2.64  & 21.19         \\
Initializer+DU     & 3.16 & 2.76  & 2.71  & 2.54  & 2.51  & 4.22         \\
\cmidrule{2-7}
PtychoDV         & 5.45  & 4.73  & 4.54  & 4.63  & 4.56  & 17.08 \\
\bottomrule
\end{tabular}
\end{center}
}
\end{table}

\begin{figure}
  \centering
  \includegraphics[width=.425\textwidth]{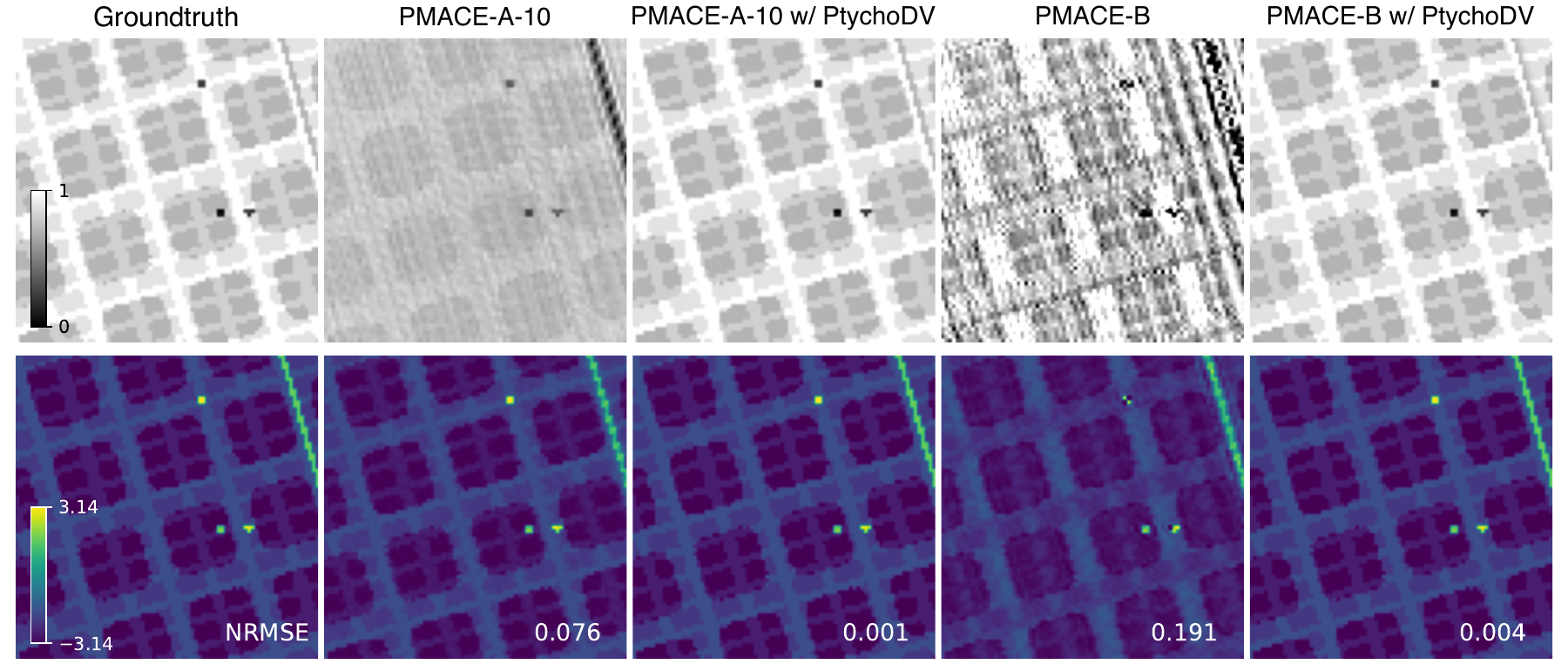}
  \caption{Visual results of PMACE tested on noise-free data generated using different probe and different initialization. The magnitude and the phase of the reconstructed images are shown in the top and the bottom row, respectively. NRMSE values of each method is labeled in the right bottom of each image. This figure shows that PMACE with a small number of iterations can achieve better performance by using PtychoDV initialization than that without it. This figure also highlights that PtychoDV could also be used to compute initialization even when the testing probe is different from the probe used in training.}
  \label{fig:initializer-noiseless}
\end{figure}

\begin{table*}[]
  \scriptsize
  \caption{{Quantitative evaluation of PtychoDV and baseline methods with format of $A\pm B$ on a new noisy testing dataset generated from probe C, where $A$ and $B$ denote mean of normalized root mean-square-error (NRMSE) and standard deviation of NRMSE, respectively. The results with {\bf \color{red}the best} mean NRMSE are highlighted. Note that probe C is exclusively used for testing. This table highlights that PtychoDV can offer a reliable initialization for PMACE, even when the testing dataset is generated using a more dissimilar asymmetry probe.}}
  \label{tab:noisy-dissimilar-probe}
  \centering
  \renewcommand\arraystretch{1.1}
  \setlength{\tabcolsep}{5pt}
  {
  \begin{threeparttable}
    \begin{tabular}{llllll}
      \toprule
      Sampling pattern & $256$:$5$   & $121$:$8$  & $64$:$11$  & $25$:$19$  & $16$:$27$  \\
      \cmidrule{2-6}
        Ours-C               & 0.411 ± 1.16  & 0.409 ± 1.15  & 0.413 ± 1.12  & 0.417 ± 1.05  & 0.433 ± 1.05  \\
        PMACE-C w/o PtychoDV & 0.438 ± 1.16  & 0.423 ± 1.03  & 0.419 ± 0.77  & 0.467 ± 0.87  & 0.483 ± 0.96  \\
        PMACE-C w/ PtychoDV  & {\bf \color{red}0.394 ± 1.45}  & {\bf \color{red}0.389 ± 1.40}  & {\bf \color{red}0.389 ± 1.35}  & {\bf \color{red}0.392 ± 1.37}  & {\bf \color{red}0.420 ± 1.11}  \\
      \bottomrule
      \end{tabular}
  \end{threeparttable}}
\end{table*}

\begin{table*}[]
  \scriptsize
  \caption{{Quantitative comparison between the proposed loss function and its constituent parts with format of $A\pm B$ on testing noisy measurements, where $A$ and $B$ denote mean of normalized root mean-square-error (NRMSE) and standard deviation of NRMSE, respectively. The results with {\bf \color{red}the best} mean NRMSE are highlighted. This table shows that the proposed loss function can gain superior performance over its constituent variants.}}
  \label{tab:loss-ablation}
  \centering
  \renewcommand\arraystretch{1.1}
  \setlength{\tabcolsep}{5pt}
  {
  \begin{threeparttable}
    \begin{tabular}{llllll}
      \toprule
      Sampling pattern & $256$:$5$   & $121$:$8$  & $64$:$11$  & $25$:$19$  & $16$:$27$  \\
      \cmidrule{2-6}
        PtychoDV w/o \emph{image-wise loss} & 0.441 ± 0.59  & 0.442 ± 0.59 & 0.443 ± 0.59 & 0.447 ± 0.59 & 0.450 ± 0.59 \\
        PtychoDV w/o \emph{patch-wise loss} & 0.047 ± 0.22  & 0.055 ± 0.26  & 0.072 ± 0.32  & 0.111 ± 0.38  & 0.142 ± 0.43  \\
        PtychoDV                & {\bf \color{red}0.043 ± 0.19}  & {\bf \color{red}0.050 ± 0.23}  & {\bf \color{red}0.065 ± 0.32}  & {\bf \color{red}0.098 ± 0.36}  & {\bf \color{red}0.127 ± 0.45}  \\
      \bottomrule
      \end{tabular}
  \end{threeparttable}
  }
\end{table*}

\begin{figure}[t!]
    \centering
    \includegraphics[width=.475\textwidth]{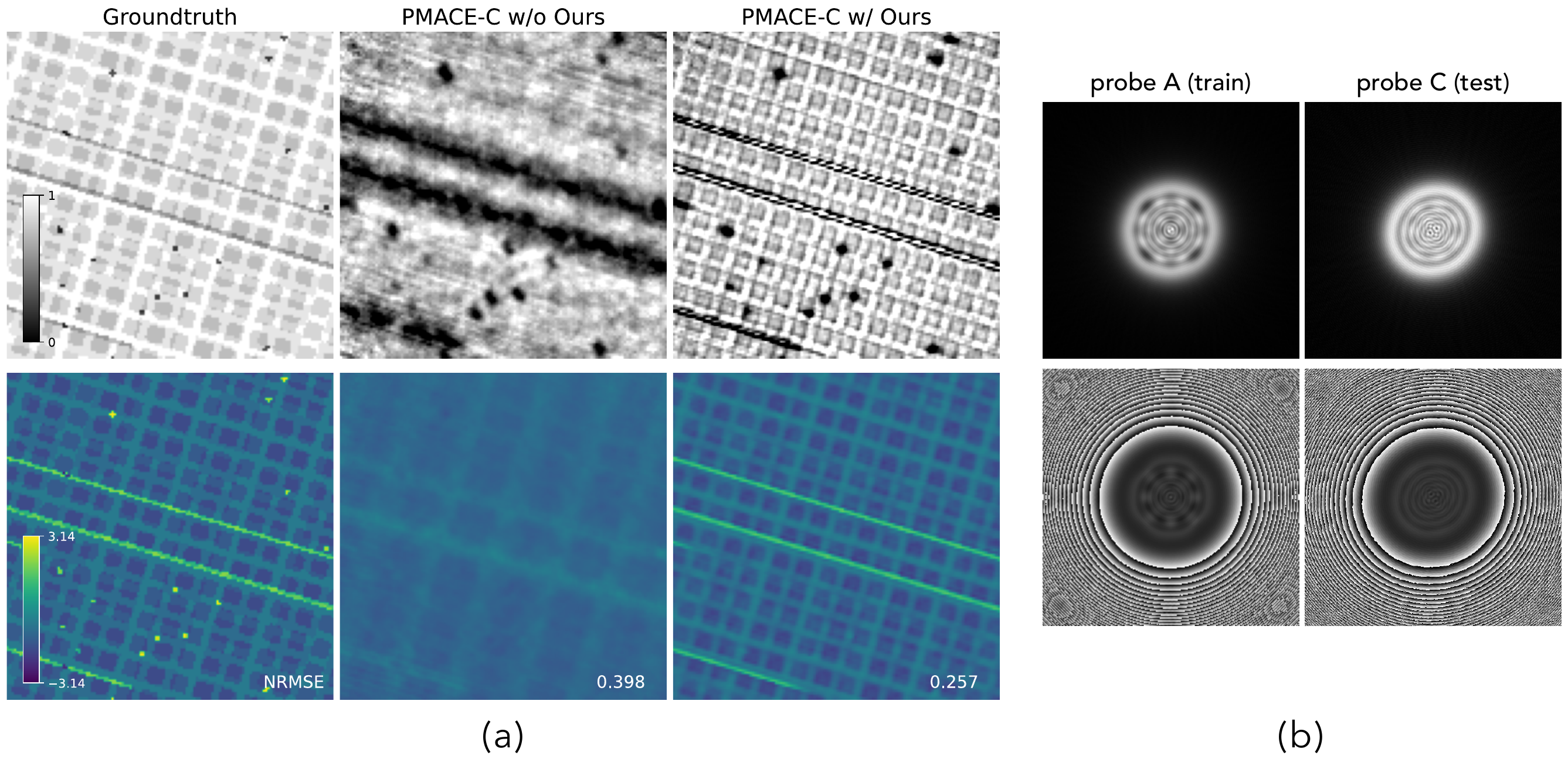}
    \caption{{\textbf{(a)} Illustration of reconstructed results obtained by PMACE with and without PtychoDV providing initialization. Experiments were conducted on a new testing dataset generated from a new probe C with a sampling pattern of $64:11$. \textbf{(b)} Illustrations of the training probe A and the new testing probe C. Probe A is used for generating the training dataset, while probe C is exclusively for testing. Note that probe A is symmetrical, whereas probe C is asymmetrical.}}
    \label{fig:noisy-dissimilar-probe}
\end{figure}

\begin{figure}
    \centering
    \includegraphics[width=.425\textwidth]{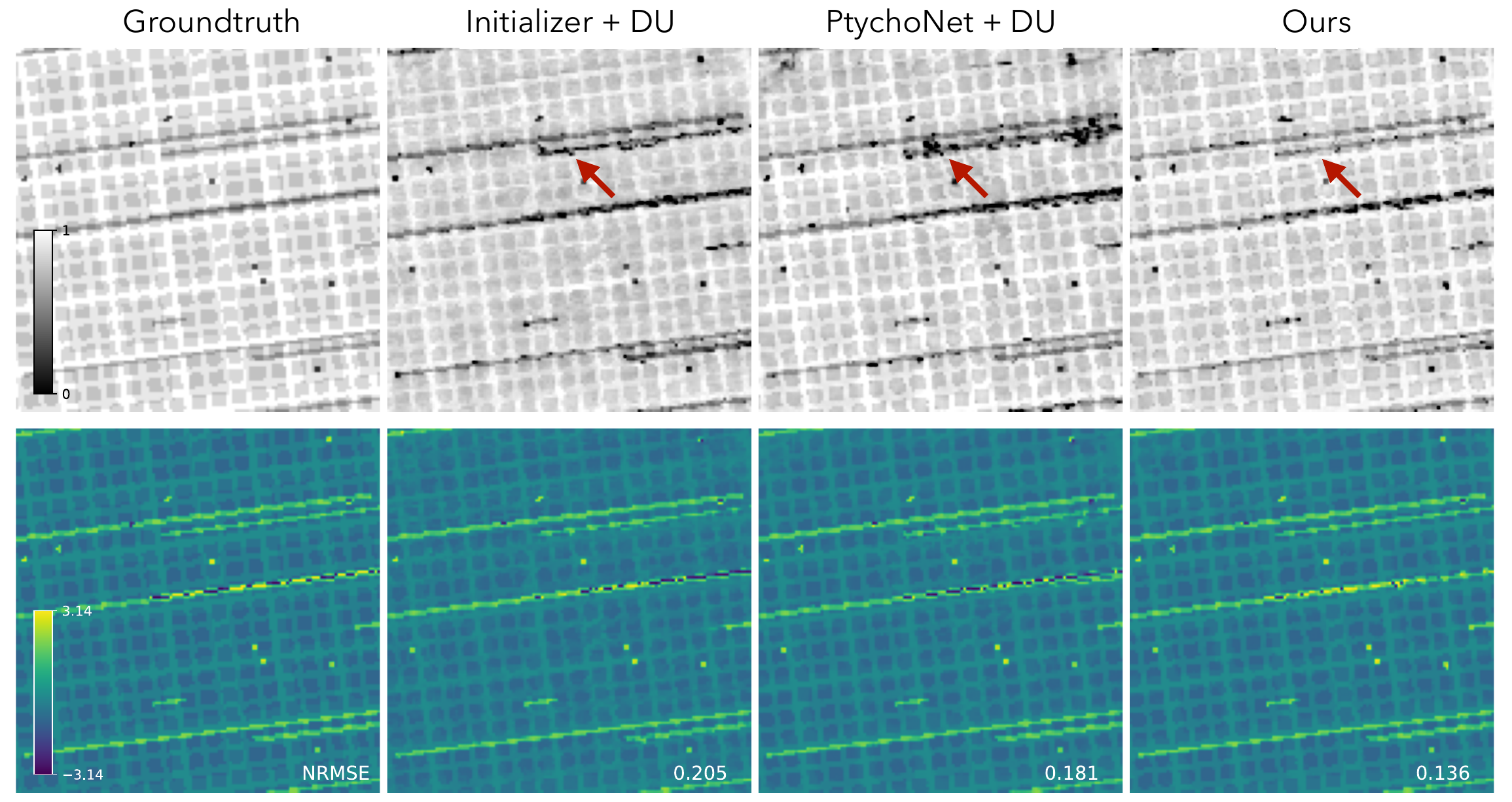}
    \caption{{Visual results of PtychoDV and other baseline methods on noisy testing data with sampling pattern of $64:11$. The magnitude and the phase of the reconstructed images are shown in the top and the bottom row, respectively. NRMSE values are included in the right bottom of each image. This figure demonstrates that PtychoDV can provide reconstructions that are more consistent with the ground truth, as highlighted by image features indicated by the red arrow.}}
    \label{fig:du_variants}
\end{figure}

{
\appendix
\section{Appendix}
This appendix reports experimental results for noise-free cases. We synthesized noise-free measurements as $\ybm_i^2 {=} |\Fbm\Pbm\xbm_i|^2$. The other experimental setups are identical to those described in Sec.~\ref{sec:exp-setup}.
Table~\ref{tab:exp-noiseless} summarizes the same type of quantitative evaluation and testing time as Table~\ref{tab:exp-noisy}, but on noise-free testing data, corroborating the same conclusions drawn from Table~\ref{tab:exp-noisy}.
Table~\ref{tab:exp-initializer-noiseless} provide a quantitative evaluation and testing time for PtychoDV and PMACE on noise-free testing data, stimulated with different probes and different sampling patterns.
Figure~\ref{fig:initializer-noiseless} shows visual results of PMACE on noise-free data with and without initialization generated by PtychoDV. This figure demonstrates that PMACE initialized by PtychoDV can provide results more consistent with the ground truth than those without it.

{Table~\ref{tab:gpu-usage} shows memory usage of PtychoDV and other baseline methods, demonstrating that ViT-based methods exhibit lower memory complexity. We attribute this to the smaller dimensions of 1D latent features in ViT compared to the 2D feature maps in CNN.

We also validated PtychoDV on testing dataset generated from an asymmetry \emph{probe C}. This new \emph{probe C} is more dissimilar to \emph{probe A} compared to \emph{probe B}. We tested the application of PtychoDV for providing a reliable initialization for PMACE on this new dataset. Table~\ref{tab:noisy-dissimilar-probe} and Figure~\ref{fig:noisy-dissimilar-probe} show quantitative and visual results on new testing dataset, respectively. Both Table~\ref{tab:noisy-dissimilar-probe} and Figure~\ref{fig:noisy-dissimilar-probe} demonstrate that PtychoDV can provide a reliable initialization for PMACE even when the testing dataset is generated using a more dissimilar asymmetry probe.

We performed experiments comparing the proposed loss function and its constituent parts. We summarized the quantitative results in Table~\ref{tab:loss-ablation}. Table~\ref{tab:loss-ablation} shows that the proposed loss function can provide superior performance over its constituent variants.

Figure~\ref{fig:du_variants} illustrates the reconstructed images of PtychoNet+DU, Initilizer+DU and PtychoDV. Figure~\ref{fig:du_variants} shows that PtychoDV can provide reconstructions that are more consistent with the ground truth, as highlighted by image features indicated by the red arrow. 
}
}

\end{document}